# RECONSTRUCTION OF TOF SPECTROMETER EXPERIMENTAL DATA USING THE STEP-BY-STEP SHIFT METHOD

**A. V. Novikov-Borodin**

*Institute for Nuclear Research (INR), Russian Academy of Sciences,*
*Moscow, 117312 Russia*
*e-mail: novikov.borodin@gmail.com*

**Abstract**—A method for mathematical treatment is considered for experimental data from pulsed time-of-flight spectrometers, whose response to measurement-initiating pulses is represented by their convolution with the pulse response of the system. The proposed method of step-by-step shifting and its modifications are intended to correct the system response when the initiating pulse is distorted and also to reconstruct the response from a shorter pulse. Within certain physical constraints, this method is an alternative for technical methods associated with difficult scientific and technical problems that require considerable material costs. The inaccuracies of the reconstruction are estimated. An example of the reconstruction of experimental data from the neutron time-of-flight spectrometer at the INR is considered. The method is applicable to reconstruction of output data from any systems that are linear with respect to the input signals and are invariant to time shifts.

---

# РЕКОНСТРУКЦИЯ ЭКСПЕРИМЕНТАЛЬНЫХ ДАННЫХ ВРЕМЯПРОЛЁТНЫХ СПЕКТРОМЕТРОВ МЕТОДОМ ПОШАГОВОГО СДВИГА

**А. В. Новиков-Бородин**

*Институт ядерных исследований РАН, Москва, Россия*
*e-mail: novikov.borodin@gmail.com*

**Аннотация.** Предлагается и анализируется метод математической обработки результатов измерений на импульсных времяпролётных нейтронных спектрометрах, позволяющий реконструировать экспериментальные данные, устранив из них искажения при различных нестабильностях и изменениях инициирующего импульса во время сбора информации. Предлагаемый метод пошагового сдвига и его модификации можно использовать для реконструкции отклика спектрометра на импульс меньшей длительности, что в определённых рамках является альтернативой техническим методам, сопряжённым с научно-техническими проблемами и большими материальными затратами. Оценивается погрешность реконструкции. Рассматривается реконструкция экспериментальных данных, полученных на времяпролётном нейтронном спектрометре ИЯИ РАН.

# INTRODUCTION

One of the most important characteristics of pulsed neutron time-of-flight spectrometers is the duration of a pulse of accelerated particles that initiates a neutron flux with different energies used in time-of-flight (TOF) measurements. An increase in the duration of the initiating pulse leads to an energy spread of the neutrons arriving at the target, which directly affects the spectrometer resolution. Thus, a change in the shape of the initiating pulse during measurements causes broadening of spectral lines of the investigated substances. The presence of after-pulses leads to the occurrence of shifted spectral lines, which can be perceived by the experimenter as additional resonances if the change in the initiating pulse is not taken into account and, therefore, lead to erroneous interpretation of the results of the experiment as a whole.

Elimination of distortions of the initiating pulse and reduction of its duration help to solve the problem radically; however, often this task is associated with a number of scientific and technical problems and requires considerable material costs. In this connection, the development of mathematical methods that allow one not only to reconstruct real spectra of investigated substances, but also to increase the spectrometer resolution, is a priority task. From the mathematical standpoint, such reconstruction problems belong to the class of incorrectly posed problems [1]. Nevertheless, the problem of reconstruction and optimization can be solved for neutron time-of-flight spectrometers, such as TRONS (Troitsk Neutron Spectrometer) at the INR [2].

In this paper, the so-called step-by-step shift method is proposed for the mathematical reconstruction and optimization of the spectrometer response and its modifications are considered.

## 1. STATEMENT OF THE RECONSTRUCTION PROBLEM

In pulsed neutron TOF spectrometers, a neutron flux is initiated by a short pulse of charged particles (usually protons or electrons) that interact with a solid target made of high-$Z$ elements. Generated neutrons decelerate in the moderator and are separated by energy based on their time of flight in lengthy neutron channels. The beam instability during data acquisition leads to various distortions in the shape of initiating pulses, to the occurrence of background noises, after- pulses, etc., which significantly degrade the spectrometer resolution.

The results of two series of independent measurements on a $^{181}Ta$ target at the TRONS setup is presented in Fig. 1. The numbers of the timing channels of the diagnostic equipment with a resolution of 150 ns/channel are plotted on the horizontal axis; the vertical axis shows the number of events in these channels (detection of γ rays from (n, γ) reactions of neutron interactions with target nuclei using an NaI-based sensor in the horizontal channel with a flight path of 50 m). The initial channel numbers in the left part of the graph correspond to interactions of cascade neutrons and γ rays with the target, whose intensity can be considered proportional to the current of the proton beam that hits the RADEX target in a set of measurements and initiates the spectrometer response. The number of events in these channels is two or three orders of magnitude greater than the number of events from delayed neutrons and the initiating pulse $S(t)$ can be identified with a good accuracy. According to Fig. 1, the shape of the initiating signal in different sets of measurements affects the spectrometer response $H^+(t)$ and the detected spectral lines $H(t)$. Thus, the after-pulses of the initiating signal in the first set of measurements do not have distinct peaks; however, their presence causes broadening of spectral lines. In this case, closely spaced peaks can merge and the spectrometer resolution decreases. The presence of a clearly defined peak of the after-pulse for the initiating signal in the second set of measurements leads to the occurrence of additional spectral lines that are not characteristic of analyzed samples upon a single pulse. This can result in erroneous interpretation of experimental results.

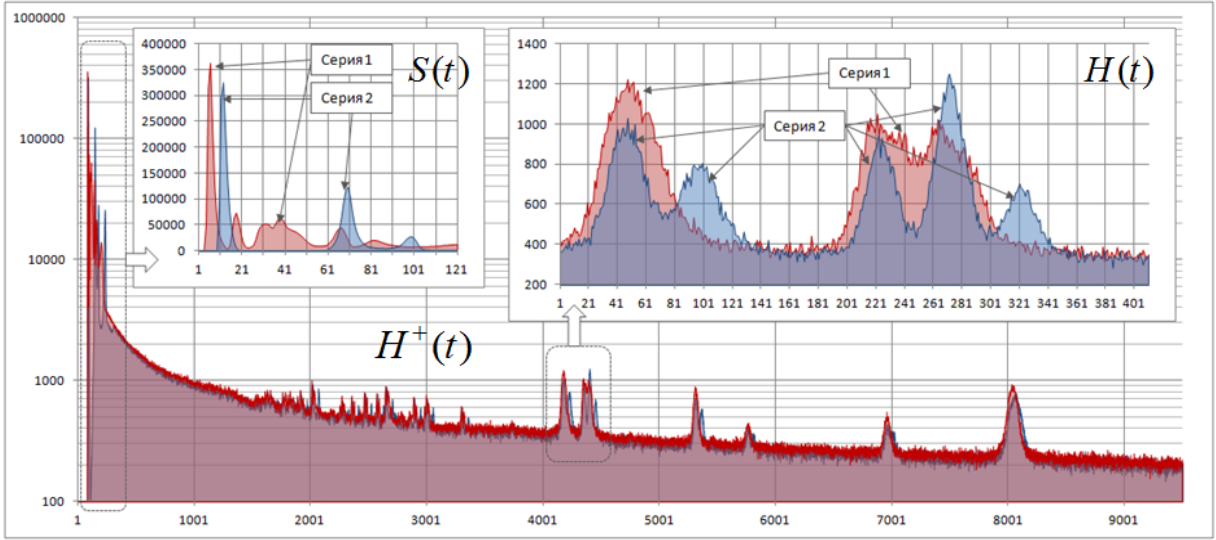

**Fig. 1.** The results of two series of independent measurements of $H^+(t)$ at the TRONS setup with different forms of the initiating signal $S(t)$. The numbers of time channels of the diagnostic equipment are on the horizontal axis; the numbers of events in them are on the vertical axis.

The total spectrometer response $H^+(t)$ to the initiating signal, which is described by the source function $S(t)$, can be represented as an integral equation:

$$H^+(t) = \lambda S(t) + \int_T S(t-x)h(x)dx = \lambda S(t) + S*h, \tag{1}$$

where $T$ is the time interval on which the measurements are taken, $\lambda$ is a certain coefficient that relates the source to the direct response of the system, $h(t)$ is the pulse response of the spectrometer to an ideal source corresponding to the Dirac delta function $\delta(0)$, and $S*h$ is the convolution of the $S$ and $h$ functions [1, 3].

Essentially, the $H(t)$ response of the spectrometer minus the source function:

$$H(t) = H^+(t) - \lambda S(t) = S*h = \int_T S(t-x)h(x)dx \tag{2}$$

is a Fredholm integral equation of the first kind [1] with a kernel corresponding to the source function, which, generally speaking, differs in different sets of measurements.

The total (1) and net (2) spectrometer responses are linear with respect to the source function and have the following properties:

(i) If the source function has the form $S(t) = \alpha_1 s_1(t) + \alpha_2 s_2(t) + ...$, where $\alpha_1, \alpha_2, ...$ are any real numbers, the corresponding response of the spectrometer is

$$H(t) = \alpha_1 h_1(t) + \alpha_2 h_2(t) + ..., \tag{3}$$

where $h_1(t), h_2(t), ...$ are spectrometer responses to source functions $s_1(t), s_2(t), ...$;

(ii) From Eq. (3), in particular, it follows that, if $s_i(t) = s(t - t_i)$, where $t_1, t_2, ...$ are any real numbers, then $h_i(t) = h(t - t_i)$ and the source function $S(t) = \alpha_1 s(t - t_1) + \alpha_2 s(t - t_2) + ...$ corresponds to the response

$$H(t) = \alpha_1 h(t - t_1) + \alpha_2 h(t - t_2) + .... \tag{4}$$

Thus, the operations of addition, multiplication by a number, and shift are admissible transformations for responses $H^+(t)$ and $H(t)$.

The problem of reconstructing the total spectrometer response $H^+(t)$ is reduced to the search for combinations of these operations on the $H^+(t)$ response for which the source function $S(t)$ entering it takes the required form $S_{opt}(t)$. In this case, due to the permissibility of operations, the response function itself will also be optimized:

$$H^+(S(t),t) \mapsto H^+_{opt}(S_{opt}(t),t). \tag{5a}$$

The source function $S(t)$ and the corresponding net response of the spectrometer $H(t)$ are formally separated from each other; however, they are nevertheless connected according to Eq. (2). The problem of reconstructing the net response $H(t)$ is reduced to the search for combinations of the operations of addition, multiplication by number, and shifting for the source function $S(t)$, which optimizes its form to obtain $S_{opt}(t)$, while similar operations on the response $H(t)$ optimize it to $H_{opt}(t)$:

$$S(t) \mapsto S_{opt}(t), \quad H(t) \mapsto H_{opt}(t). \tag{5b}$$

In many practical cases, the separate optimization of the source function and the net response according to Eq. (5b) substantially simplifies the mathematical treatment, allowing the reconstruction of various parts of the response $H(t)$.

## 2. THE RECONSTRUCTION PROBLEM FOR DISCRETE FUNCTIONS

In practice, the spectrometer response is a discrete function, whose timing is determined by the width of the channel of the diagnostic equipment, and the value of the function is equal to the number of detected events in each interval. For the data shown in Fig. 1, the total number of time channels of the diagnostic equipment was 149 000 with a resolution of 150 ns/channel, i.e., the total measurement interval $T$ reached 22.35 ms with a time sampling $\Delta t = 150$ ns.

If the $k$th time channel is denoted $\delta_k$ and unity is assigned to it, any function $F(t)$ in the interval $T$ can be represented in a discrete form:

$$F(t) \cong F = \sum_k f_k \delta_k, \tag{6}$$

where $f_k$ is the value of the function $F(t)$ on the $k$th time interval. In fact, $\delta_k$ is a Kronecker symbol and is defined on the minimum-duration interval.

A linear combination of functions $aF(t) + bG(t)$ can be written as

$$aF(t) + bG(t) \cong aF + bG = \sum_k (af_k + bg_k)\delta_k, \tag{7}$$

and the shift of the function $F(t)$ for the interval $t_0 = m\Delta t$ is described by the expression

$$F(t - t_0) \cong F_m = \sum_k f_k \delta_{k+m}. \tag{8}$$

In accordance with Eqs. (6)–(8), the spectrometer responses $H^+ \cong H^+(t)$ and $H \cong H(t)$ can be presented as:

$$H^+ = \lambda \sum_k s_k \delta_k + \sum_m \sum_k s_k h_m \delta_{k+m}, \quad H = \sum_m \sum_k s_k h_m \delta_{k+m}, \tag{9}$$

where $S(t) \cong S = \sum_k s_k \delta_k$ is the discrete representation of the source function and $h(t) \cong \sum_m h_m \delta_m$ is the pulse response of the spectrometer to the source $S=\delta_0$.

The problem of reconstructing the spectrometer response in the discrete representation (9) is similar to Eqs. (5a) and (5b).

## 3. THE METHOD OF STEP-BY-STEP SHIFTING

Let the source function $S$ and the net response $H$ to it be represented in accordance with (9) in the following form:

$$S = \sum_{k=0}^{K} s_k \delta_k, \quad H = \sum_{k=0}^{K} \sum_{m=M_0}^{M} s_k h_m \delta_{k+m}. \tag{10}$$

The function $S$ is defined on the interval $[0, K]$, and $H$ is defined on the interval $[M_0, M + K]$.

Assuming that $S_{opt} = s_0 \delta_0$, the reconstruction problem (5b) consists in determining the response function $H_{opt} = s_0 \sum_{m=M_0}^{M} h_m \delta_m$, which is defined on the interval $[M_0, M]$.

The step-by-step shift method consists in successively zeroing the coefficients $s_1, s_2, \ldots$ in Eq. (10) using the permissible transformations (7) and (8) over the function $S$; therefore, $S \mapsto S_{opt} = s_0 \delta_0$, and, according to Eq. (5b), $H \mapsto H_{opt}$. This algorithm is performed using the iterative equations:

$$\begin{bmatrix} S^1 = S^0 - \dfrac{s_1^0}{s_0^0} S_1^0, \\ S^2 = S^1 - \dfrac{s_2^1}{s_0^1} S_2^1, \\ \ldots \\ S^{n+1} = S^n - \dfrac{s_{n+1}^n}{s_0^n} S_{n+1}^n. \end{bmatrix} \quad \begin{bmatrix} H^1 = H^0 - \dfrac{s_1^0}{s_0^0} H_1^0, \\ H^2 = H^1 - \dfrac{s_2^1}{s_0^1} H_2^1, \\ \ldots \\ H^{n+1} = H^n - \dfrac{s_{n+1}^n}{s_0^n} H_{n+1}^n. \end{bmatrix}, \tag{11}$$

where $S^0 = S$, and $H^0 = H$, while $S^n = \sum_{k=0}^{K+n} s_k^n \delta_k$ and $H^n = \sum_{k=0}^{K+n} \sum_{j=m}^{M} s_k^n h_j^n \delta_k$, are, respectively, the source function and the response to it at the $n$th iteration step. Upon successive zeroing in the source function $S^n$, the coefficients are $s_1^n = s_2^n = \cdots = s_n^n = 0$; therefore:

$$S^n = s_0 \delta_0 + \sum_{k=n+1}^{K+n} s_k^n \delta_k, \quad H^n = s_0 \sum_{m=M_0}^{M} h_m \delta_m + \sum_{k=n+1}^{K+n} \sum_{m=M_0}^{M} s_k^n h_m^n \delta_{m+k}. \tag{12}$$

If the number of iterations is $N > K+M-M_0$, the second terms in Eq. (12) go beyond the considered domain of the definition of functions, $S$ and $H$: $S^N \mapsto S_{opt} = s_0 \delta_0$ in the interval $[0, K]$; while, in the interval $[M_0, K+M]$, $H^N \mapsto H_{opt} = s_0 \sum_{m=M_0}^{M} h_m \delta_m$, which means finding the solution to the reconstruction problem.

Figure 2 illustrates the process of reconstruction of the response $H$ using the step-by-step shift method for the source $S = 1.5 \cdot \delta_{0-3}+1.0 \cdot \delta_{4-7}+0.6 \cdot \delta_{8-11}+0.3 \cdot \delta_{16-19}+0.6 \cdot \delta_{20-23}$, where $\delta_{i-i+k} = \delta_i + \delta_{i+1} + \ldots + \delta_{i+k}$. The discreteness of the source function was four minimum time intervals; thus, the iteration step was also four intervals. The vertical dotted–dashed lines indicate the initial boundaries of the source function ($K \approx 30$) and the response ($K + M - M_0 \approx 60$). In these intervals, the source function already takes an optimal form at the eighth iteration step ($8 \cdot 4 = 32 > K \approx 30$), and the response takes its optimal form at the 16th step ($16 \cdot 4 = 64 > K + M - M_0 \approx 60$). Thus, in this example, N = 16 is the minimum number of iterations necessary for the reconstruction.

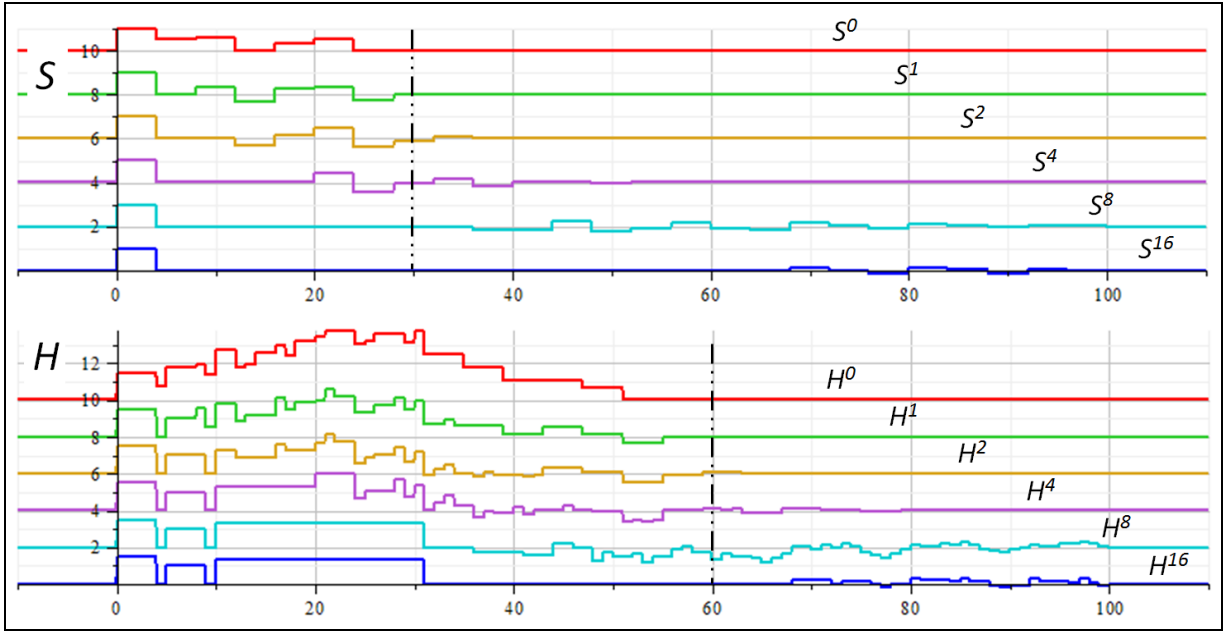

**Fig. 2.** The diagrams of the reconstruction of the source function $S$ and the system response $H$ using the method of step-by-step shift along the time channels (horizontal axis) at various iteration steps.

In practice, the number of time channels can be much larger than in this example. Thus, for example, as mentioned above, the total number of time channels in the response shown in Fig. 1 was 149000. In spite of the fact that the reconstruction method (5b) makes it possible to consider the response in parts, a great number of iterations may be required. In many cases, the step-by-step shift method can be modified, significantly reducing the number of iterations required for the reconstruction.

## 4. THE MODIFIED METHOD

Let the source function $S$ and the net response $H$ to it be represented in the form:

$$S = s_0\delta_0 + s_k\delta_k, \; H = s_0 \sum_{m=M_0}^{M} h_m\delta_m + s_k \sum_{m=M_0}^{M} h_m\delta_{k+m}, \qquad (13)$$

i.e., the source function $S$ constitutes two time-separated initiating pulses: the main one with the amplitude $s_0$ and the after-pulse with the amplitude $s_k$. The coefficients of the source function are $s_1 = s_2 = \cdots = s_{k-1} = 0$; therefore, the first $k - 1$ steps of the iterative process (11) do not change either the source function or the response and can be omitted. The first change of functions occurs at the $k$th step of the iteration (11); in this case, the coefficient at $\delta_k$ of function $S$ is set to zero and its coefficient at $\delta_{2k}$ becomes nonzero. The next $k - 1$ steps of the iterative process (11) also do not change the form of the functions. The functions will change on the $k$th, $2k$th, $4k$th, etc. step of the iteration (11). Omitting steps in Eq. (11) that do not change the $S$ and $H$ functions and denoting $a = s_k/s_0$, we obtain the following iteration equations:

$$S^{n+1} = S^n + a^{2^n} S^n_{2^n k}, \; H^{n+1} = H^n + a^{2^n} H^n_{2^n k}, \qquad (14)$$

where $S^0 = S$, $S^1 = S^0 - aS^0_k$, $H^0 = H$, $H^1 = H^0 - aH^0_k$.

The modified method significantly reduces the number of iteration steps needed for the reconstruction. Thus, if $N$ is the number of iterations required for the reconstruction using the step-by-step shift method, the number of iterations $N^*$ in the modified method is: $N^* = \log_2(N/k)$. As an example, even for $k = 1$, instead of the 64 iterations required for the reconstruction using the step-by-step shift method, only 6 iterations will be required if the modified method is used; only 7 iterations will be needed instead of 128; and only 8 iterations will be instead of 256.

In addition, for $a < 1$, the after-pulse amplitude at the $n$th iteration step decreases by a factor of $a^{2n}$ and, according to Eq. (14), the distortion of the response to the main pulse also decreases. Table 1 shows the distortion of the responses to the main pulse in percent at the relation for the pulse amplitudes $a < 1$ and n iterations in the modified method. Thus, at $a = 0.3$, the error from the after-pulse is 30%; nevertheless, after the second iteration, it is only 0.8%, and, if such an error is suitable, the reconstruction can be considered complete.

Table 1. The distortions [%] of the response to the main pulse with the amplitude a for n iterations

| $a \setminus n$ | 0 | 1 | 2 | 3 | 4 | 5 |
|---|---|---|---|---|---|---|
| 0.9 | 90 | 81 | 65.61 | 43.05 | 18.53 | 3.434 |
| 0.7 | 70 | 49 | 24.01 | 5.77 | 0.332 | 0.001 |
| 0.5 | 50 | 25 | 6.25 | 0.391 | 0.002 | $2.33 \cdot 10^{-8}$ |
| 0.3 | 30 | 9 | 0.81 | $6.56 \cdot 10^{-3}$ | $4.30 \cdot 10^{-7}$ | $1.85 \cdot 10^{-15}$ |
| 0.1 | 10 | 1 | $10^{-2}$ | $10^{-6}$ | $10^{-14}$ | $10^{-30}$ |

The reconstruction of the response to the source function $S = \delta_{0 \div 3} + 0.85 \cdot \delta_{8 \div 11}$ using the modified method is diagrammatically shown in Fig. 3. For three iterations, the distortions in the system response $H^3$ go beyond the boundaries of the considered region; i.e., three iterations are enough to reconstruct the response.

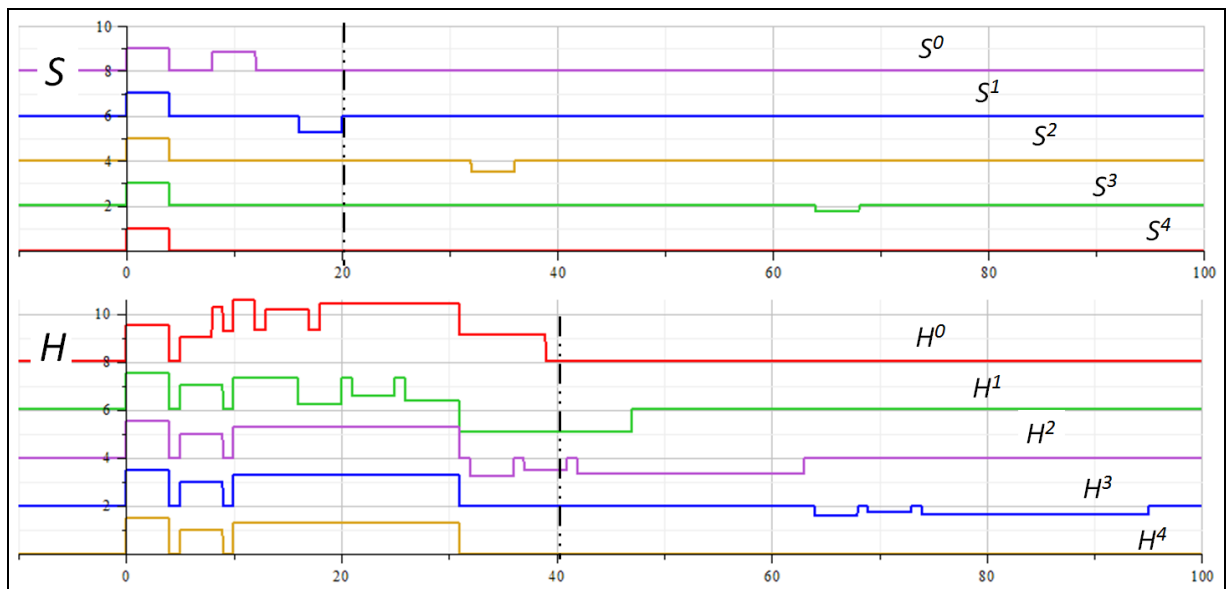

**Fig. 3.** The diagrams of the reconstruction of the source function $S$ and the system response $H$ using the modified method at various iteration steps.

In the case of $a = 1$, there is no reduction in the amplitude of the distortion and the reconstruction is carried out only by leading distortions outside the detection domain.

In the case where $a > 1$, the application of both the step-by-step shift method and the modified method will lead to an exponential increase in the distortions and to the divergence of the iterative sequence during the reconstruction. The problem can be solved if we simulate the pulse $S$ from Eq. (13) in the form

$$S = s_0 \delta_0 + \sum_p \frac{s_k}{p} \delta_k , \qquad (15)$$

i.e., by representing the after-pulse $s_k \delta_k$ as the sum of $p$ after-pulses with an amplitude $s_k/p$. The $p$ value is selected such that $a = s_k/(p s_0) \leq 1$. The reconstruction problem can be solved by successively eliminating the effect of *p* after-pulses with amplitude $s_k/p$ using the modified method.

The modified reconstruction method (14) and the pulse-simulation method (15) allow finding a solution to the reconstruction problem for an arbitrary number of after-pulses with an arbitrary amplitude

## 5. IMPROVING THE RESOLUTION

The possibility of eliminating the influence of selected after-pulses or parts of the initiating pulse is an important advantage of the modified method. This possibility can be used to reconstruct the spectrometer response to the initiating pulse with a shorter duration or with shorter edges, which will correspond to an increase in the resolution of the spectrometer exceeding its design parameters up to the physical constraints.

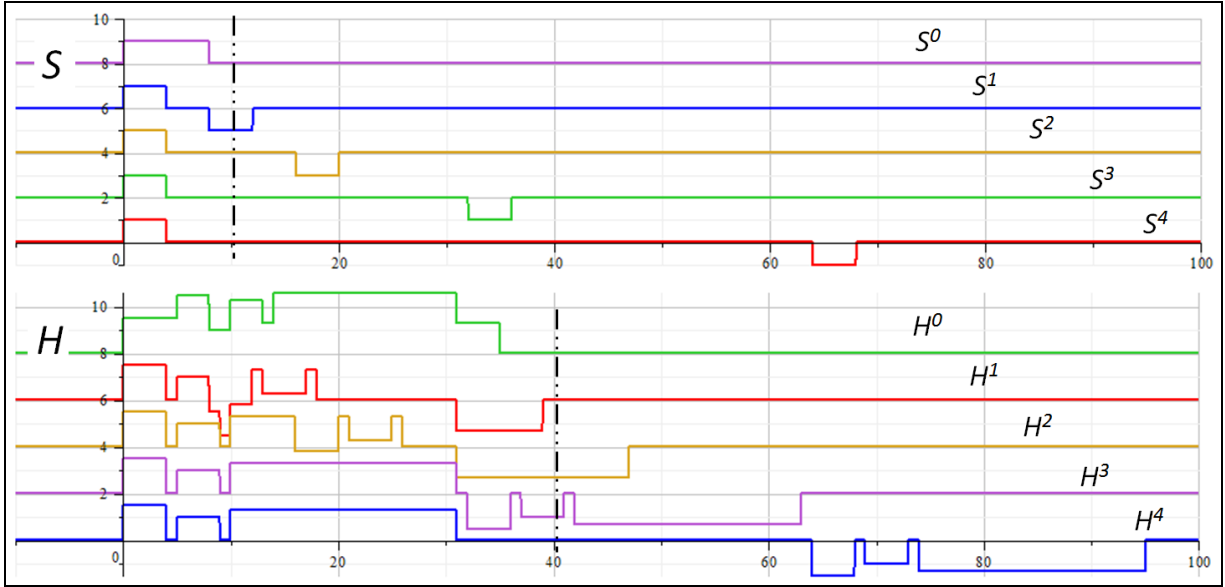

**Fig. 4.** The diagrams for reconstructing the system response to a shorter-duration pulse using the modified method at various iteration steps.

Let the source function *S* and the net response *H* to it be represented in the form:

$$S = s_0\delta_0 + s_0\delta_1, \quad H = s_0 \sum_{m=M_0}^{M} h_m \delta_m + s_0 \sum_{m=M_0}^{M} h_m \delta_{m+1}, \quad (16)$$

i.e., the source function *S* is composed of two successive initiating pulses with the same amplitude $s_0$.

Equation (16) is a special case of Eq. (13) for $\delta_k = \delta_1$ and $s_k = s_0$, and the modified reconstruction method is applicable to it. Essentially, the spectrometer response to the half-duration initiating signal will be obtained during reconstruction, which means an increase in the spectrometer resolution.

Figure 4 shows the reconstruction diagram for the response to a pulse with a shorter duration when the source function $S = \delta_{0 \div 7}$ is represented in the form $S = \delta_{0 \div 3} + \delta_{4 \div 7}$. Since the coefficient $a = s_1/s_0 = 1$, there is no decrease in the amplitude of the distortions and the reconstruction is carried out exclusively by leading distortions out of the considered domain of definition.

Similarly, it is possible to reconstruct the response to a pulse with shorter duration. The possibility of reducing the duration of pulses is limited by the sampling interval and the physical processes that occur during formation of a neutron flux and neutron interactions with target nuclei.

Another limiting factor is the leading edge of the initiating pulse, for which $s_0 < s_1 < \cdots < s_{max}$; thus, the coefficient $a = s_k/s_0 > 1$, which leads to the divergence of the iterative sequence. The problem can be solved using the pulse-simulation method (15); in this case, a large number of iteration sequences may be needed. Their number can be reduced using a method for simulating the pulse edges.

Thus, for the response to the source function $S = \delta_{0 \div 1} + 2\delta_{2 \div 3} + \delta_{4 \div 5}$ (see Fig. 5) to be reconstructed using Eq. (15), at least three iterative sequences are required. However, by representing this function in the form $S = \delta_{0 \div 3} + \delta_{2 \div 5}$, it is possible to reconstruct the response using one iteration sequence (Fig. 5). Since the coefficient $a = s_1/s_0 = 1$, the distortion amplitude does not decrease and the reconstruction is carried out by leading the distortions out of the considered domain of definition.

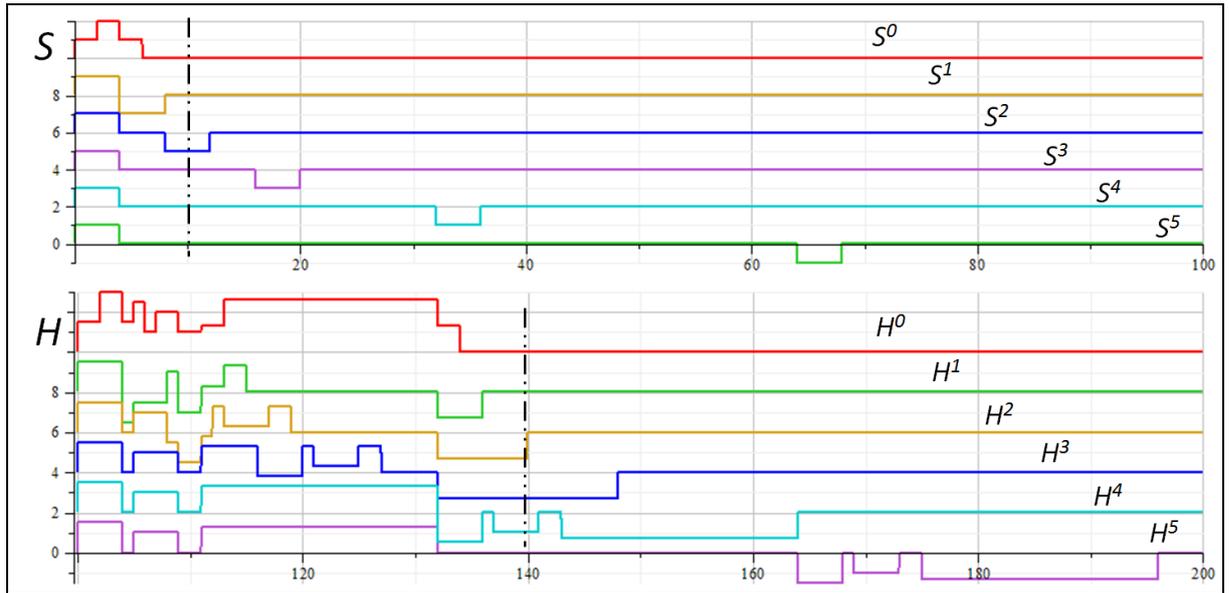

**Fig. 5.** The diagrams for reconstructing the system response to a pulse with edges using the signal-simulation method.

To reduce the number of iterations, it may be useful to change to convergent iterative sequences, e.g., by representing the considered function $S = \delta_{0 \div 3} + \delta_{2 \div 5}$ in the form $S = \delta_{0 \div 3} + \frac{1}{2}\delta_{2 \div 5} + \frac{1}{2}\delta_{2 \div 5}$. In this case, two iterative sequences are required, but, according to Table 1, they consist of only three iteration steps to achieve an accuracy of 0.4%.

Thus, a function with an edge can be reduced to a rectangular form by using the representation of the source function in the form

$$S = \sum_k a_k U_k, \qquad (17)$$

where $U_k = \delta_{k \div k+m}$ and $|a_k| \leq 1$.

In practice, if the initiating pulse has a complex form, it is possible to average it to a rectangle with a duration determined by the full width at half-maximum of the pulse. In this case, the reconstruction accuracy will be determined by this averaging.

Figure 6 illustrates the reconstruction of the spectrometer response $H(t)$ shown in Fig. 1 in series of measurements 2 in the resonance region using the modified method with the averaging of the conversion coefficient $a$ over the half-height of the initiating pulse. The reconstructed region had a duration of 400 channels, which is 150 ns · 400 = 60 μs. The conversion coefficient was selected to be $a = 0.5$; the iteration step was determined from the delay between the main pulse and the after-pulse and amounted to 6 μs.

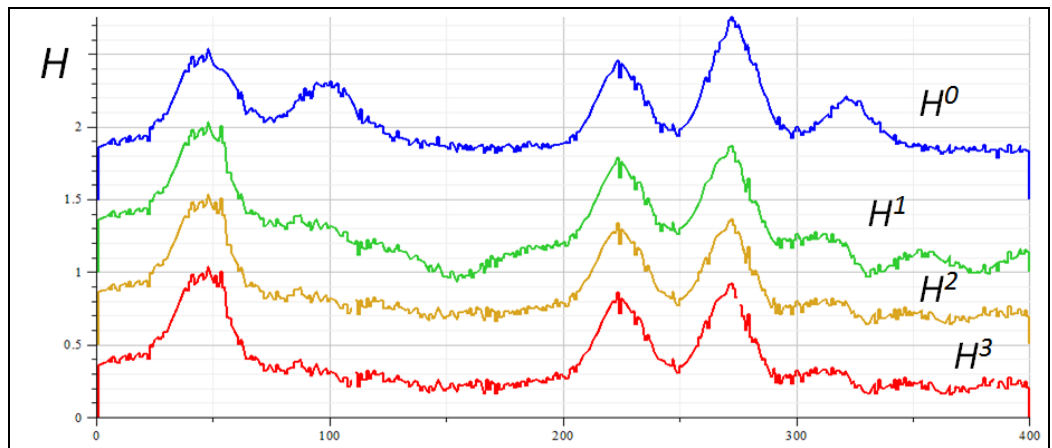

**Fig. 6.** The reconstruction of the experimental data of the TRONS spectrometer using the modified method: ($H^0$, $H^3$) initial and reconstructed responses of the system. The numbers of the time channels of the diagnostic equipment are on the horizontal axis; the numbers of events in them are on the vertical axis.

In accordance with the modified method (compare with Fig. 3), the after-pulse in the initial initiating signal $S^0$ is eliminated at the initial step of the iteration ($S^1$, $H^1$), and the respective false resonance lines in the initial response of the system $H^0$ disappear. At the second iteration step, distortions in $H^1$ that correspond to the effect of the resulting negative after-pulse in $S^1$ are eliminated. The reconstructed response $H^2$ corresponds to the optimized initiating signal $S^2$, for which the amplitude of the after-pulses is several times smaller than the amplitude of the after-pulses in the original signal. If the achieved accuracy is inadequate, the influence of the remaining after-pulses can be eliminated in a similar way.

Increasing the resolution of spectrometers with respect to the moderation time includes great technical difficulties, such as optimization of the shape of the initiating pulse, reduction of its duration, and an increase in the flight path. Thus, e.g., the flight path must be approximately doubled (in the TRONS at the INR, from 50 to 100 m) to increase the resolution of a TOF spectrometer; incidentally, the intensity of the neutron flux will drop by approximately four times. Increasing the resolution of the spectrometer using mathematical methods does not require large expenditures and is very promising.

## CONCLUSIONS

The proposed mathematical methods of step-by-step shifting and its modification are used to reconstruct the experimental data of measuring systems that are linear with respect to the input signals and invariant to time shifts, in particular, for TOF spectrometers. By using mathematical treatment of experimental data from TOF spectrometers, it is possible to eliminate distortions in them, which are caused by the instability of initiating pulses and to optimize measurements. In addition, these methods allow us to reconstruct, within certain margins, the response from shorter initiating pulses, which corresponds to an increase in the resolution of available spectrometers and is a promising alternative to technical methods that are associated with difficult scientific and technical problems and require large material and financial costs.

# ВВЕДЕНИЕ

Одной из важнейших характеристик импульсных времяпролётных нейтронных спектрометров является длительность импульса ускоренных частиц, инициирующих поток нейтронов различной энергии, используемых во времяпролётных измерениях. Увеличение длительности инициирующего импульса приводит к энергетическому разбросу нейтронов, приходящих на исследуемую мишень, что напрямую влияет на разрешающую способность спектрометра. Так, изменение формы инициирующего импульса во время измерений приведёт к расширению спектральных линий исследуемых веществ, а наличие после-импульсов – к появлению смещённых спектральных линий, которые, если изменения инициирующего импульса не будут учтены, может быть воспринято экспериментатором как дополнительные резонансы и, следовательно, приведёт к ошибочной интерпретации результатов эксперимента в целом.

Устранение искажений инициирующего импульса, уменьшение его длительности является кардинальным решением проблемы, но зачастую эта задача сопряжена с рядом научно-технических проблем и требует значительных материальных затрат и финансовых вложений. В этой связи, разработка математических методов, позволяющих не только реконструировать реальные спектры исследуемых веществ, но также повысить разрешающую способность спектрометров, является приоритетной задачей. С точки зрения математики, подобные задачи реконструкции относятся к классу некорректно поставленных задач [1], но в случае времяпролётных нейтронных спектрометров типа ТРОНС (Троицкий Нейтронный Спектрометр) ИЯИ РАН [2] и ему подобных, задача реконструкции и оптимизации может быть решена. В данной работе для математической реконструкции и оптимизации отклика спектрометра предлагается так называемый метод пошагового сдвига и рассматриваются его модификации.

## 1. ПОСТАНОВКА ЗАДАЧИ РЕКОНСТРУКЦИИ

В импульсных нейтронных спектрометрах по времени пролёта поток нейтронов инициируется коротким импульсом заряженных частиц (обычно протонов или электронов) при взаимодействии их с твердотельной мишенью из тяжёлых элементов. Образующиеся нейтроны замедляются в замедлителе и сепарируются по энергии по времени пролёта в протяжённых нейтроноводах. Нестабильности пучка во время набора статистики приводят к различным искажениям формы инициирующих импульсов, к появлению фонов, послеимпульсов и др., что значительно ухудшает разрешающую способность спектрометров.

На Рисунке 1 представлены результаты двух серий независимых измерений на мишени из Ta-181 на установке ТРОНС [2]. По горизонтальной оси отложены номера временных каналов диагностического оборудования с разрешением 150 нс/канал, по вертикальной – количество событий в них (регистрация γ-квантов от реакций (n,γ) взаимодействия нейтронов с ядрами мишени с помощью датчика на основе NaI в горизонтальном канале с пролётной базой 50 м). Начальные номера каналов в левой части графика соответствуют взаимодействию с исследуемой мишенью каскадных нейтронов и гамма-квантов, интенсивность которых можно считать пропорциональной сумме тока пучка протонов, приходящих на мишень РАДЭКС за серию измерений и инициирующих отклик спектрометра. Количество событий в этих каналах на два-три порядка превышает количество событий от замедленных нейтронов и инициирующий импульс может быть с хорошей точностью идентифицирован. Из рисунка видно, что форма инициирующего сигнала влияет на отклик спектрометра и на регистрируемые спектральные линии. Так, послеимпульсы инициирующего сигнала в первой серии измерений не имеют ярко выраженных пиков, но их наличие приводит к уширению отдельных спектральных линий и слиянию близкорасположенных, что свидетельствует о снижении разрешающей способности спектрометра. Наличие ярко выраженного пика после-импульса инициирующего сигнала во второй серии измерений приводит к появлению дополнительных спектральных линий, нехарактерных для исследуемых образцов при единичном импульсе, что может привести к ошибочной интерпретации результатов измерений.

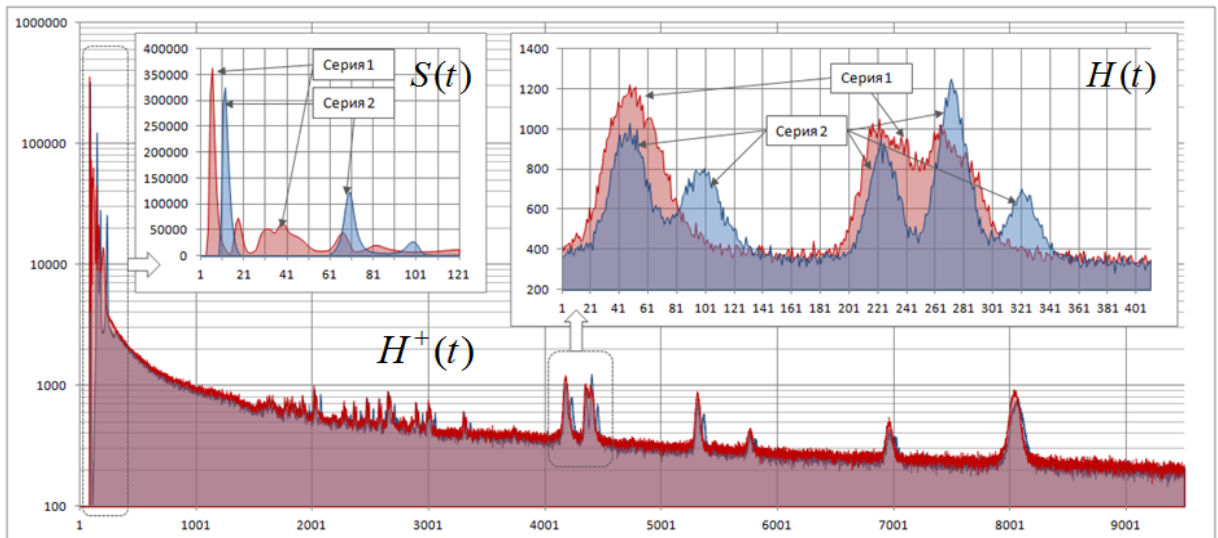

**Рисунок 1.** Результаты двух серий измерений на мишени из Ta-181.

Суммарный отклик $H^+(t)$ спектрометра на инициирующий сигнал, описываемый функцией источника $S(t)$, можно представить в виде интегрального уравнения:
$$H^+(t) = \lambda S(t) + \int_T S(t-x)h(x)dx = \lambda S(t) + S*h, \qquad (1)$$
где $T$ – временной интервал, на котором проводятся измерения, $\lambda$ – некоторый коэффициент, $h(t)$ – отклик спектрометра на идеальный источник, соответствующий дельта-функции Дирака $\delta(0)$, $S*h$ – свёртка функций $S$ и $h$.

Чистый отклик $H(t)$ спектрометра за вычетом функции источника:
$$H(t) = H^+(t) - \lambda S(t) = S*h = \int_T S(t-x)h(x)dx \qquad (2)$$
по существу является интегральным уравнением Фредгольма 1-го рода с ядром, соответствующим функции источника [1,3] и относится к классу некорректно поставленных задач.

Отклики спектрометра (1) и (2) линейны относительно функции источника, то есть:
1. Если функция источника $S(t) = \alpha_1 s_1(t) + \alpha_2 s_2(t) + ...$, где $\alpha_1, \alpha_2, ...$ – любые действительные числа, то соответствующий отклик спектрометра равен:
$$H(t) = \alpha_1 h_1(t) + \alpha_2 h_2(t) + ..., \qquad (3)$$
где $h_1(t), h_2(t), ...$ – отклики спектрометра на функции источников $s_1(t), s_2(t), ...$.

2. Из (3), в частности, следует, что если $s_i(t) = s(t-t_i)$, где $t_1, t_2, ...$ – любые действительные числа, то $h_i(t) = h(t-t_i)$ и функции источника $S(t) = \alpha_1 s(t-t_1) + \alpha_2 s(t-t_2) + ...$ будет соответствовать отклик:
$$H(t) = \alpha_1 h(t-t_1) + \alpha_2 h(t-t_2) + .... \qquad (4)$$

Таким образом, операции сдвига, сложения и умножения на число являются эквивалентными преобразованиями для откликов $H^+(t)$ и $H(t)$.

Задача реконструкции суммарного отклика спектрометра $H^+(t)$ сводится к поиску комбинаций этих операций над откликом $H^+(t)$, при которых входящая в него функция источника $S(t)$ принимает требуемый вид $S_{opt}(t)$. При этом сама функция отклика также будет оптимизирована:

$$H^+(S(t),t) \mapsto H^+_{opt}(S_{opt}(t),t). \tag{5a}$$

Задача реконструкции чистого отклика $H(t)$ сводится к поиску комбинаций операций сложения, умножения на число и сдвига над функцией источника $S(t)$, которые оптимизируют её вид до $S_{opt}(t)$, при этом аналогичные операции над откликом $H(t)$ оптимизируют его до $H_{opt}(t)$:

$$S(t) \mapsto S_{opt}(t), \quad H(t) \mapsto H_{opt}(t). \tag{5b}$$

Во многих практических случаях раздельная оптимизация функции источника $S(t)$ и отклика $H(t)$ согласно (5b) значительно упрощают реконструкцию, позволяя реконструировать различные части отклика.

Таким образом, в данной постановке задачи идентификация функции источника $S(t)$ тем или иным способом является необходимым условием оптимизации отклика.

## 2. ЗАДАЧА РЕКОНСТРУКЦИИ ДЛЯ ДИСКРЕТНЫХ ФУНКЦИЙ

На практике отклик спектрометра представляет собой дискретную по времени функцию (см. Рисунок 1). Дискретизация определяется временной шириной канала диагностического оборудования, а значение функции равно числу зарегистрированных событий в каждом интервале. Для представленных на Рисунке 1 данных общее число временных каналов диагностического оборудования составляло 149000 с разрешением 150 нс/канал, то есть общий интервал измерения $T$ составлял 22.35 мс с дискретизацией по времени $\Delta t = 150$ нс.

Если через $\delta_k$ обозначить $k$-ый временной канал и присвоить ему значение 1, то любая функция $F(t)$ на интервале $T$ может быть представлена в дискретном виде как:

$$F(t) \cong F = \sum_k f_k \delta_k, \tag{6}$$

где $f_k$ – значение функции $F(t)$ на $k$-том временном интервале $\delta_k$. Временной канал $\delta_k$ определяет шаг дискретизации и фактически является цифровым аналогом дельта-функции Дирака.

Линейная комбинация функций $aF(t) + bG(t)$ в дискретном виде может быть записана как:

$$aF(t) + bG(t) \cong aF + bG = \sum_k (af_k + bg_k)\delta_k, \tag{7}$$

а сдвиг функции $F(t)$ на интервал $t_0 = m\Delta t$ будет описываться выражением:

$$F(t - t_0) \cong F_m = \sum_k f_k \delta_{k+m}. \tag{8}$$

Согласно (6)-(8), отклики спектрометра $H^+ \cong H^+(t)$ и $H \cong H(t)$ в дискретном виде могут быть представлены как:

$$H^+ = \lambda \sum_k s_k \delta_k + \sum_m \sum_k s_k h_m \delta_{k+m}, \quad H = \sum_m \sum_k s_k h_m \delta_{k+m}, \tag{9}$$

где $S(t) \cong S = \sum_k s_k \delta_k$ – дискретное представление функции источника, а $h(t) \cong \sum_m h_m \delta_m$ – отклик спектрометра на источник $S = \delta_0$.

Постановка задачи реконструкции отклика спектрометра в дискретном представлении аналогична (5a) и (5b) для функций (9).

## 3. МЕТОД ПОШАГОВОГО СДВИГА

Пусть функция источника $S$ и чистый отклик $H$ от неё в соответствии с (9) представлены в виде:

$$S = \sum_{k=0}^{K} s_k \delta_k, \quad H = \sum_{k=0}^{K} \sum_{m=M_0}^{M} s_k h_m \delta_{k+m}. \tag{10}$$

Функция $S$ определена на интервале $[0, K]$, $H$ – на интервале $[M_0, K+M]$.

Если считать, что $S_{opt} = s_0 \delta_0$, то задача реконструкции (5b) состоит в определении функции отклика $H_{opt} = s_0 \sum_{m=M_0}^{M} h_m \delta_m$, которая определена на интервале $[M_0, M]$.

Метод пошагового сдвига заключается в последовательном обнулении коэффициентов $s_1, s_2, \ldots$ с помощью эквивалентных преобразований (7),(8) над функцией $S$, тогда на некотором рассматриваемом интервале: $S \mapsto S_{opt} = s_0 \delta_0$ и, согласно (5b): $H \mapsto H_{opt}$. Данный алгоритм реализуется с помощью итерационных уравнений:

$$\begin{bmatrix} S^1 = S^0 - \dfrac{s_1^0}{s_0^0} S_1^0, \\ S^2 = S^1 - \dfrac{s_2^1}{s_0^1} S_2^1, \\ \ldots \\ S^{n+1} = S^n - \dfrac{s_{n+1}^n}{s_0^n} S_{n+1}^n. \end{bmatrix}, \begin{bmatrix} H^1 = H^0 - \dfrac{s_1^0}{s_0^0} H_1^0, \\ H^2 = H^1 - \dfrac{s_2^1}{s_0^1} H_2^1, \\ \ldots \\ H^{n+1} = H^n - \dfrac{s_{n+1}^n}{s_0^n} H_{n+1}^n. \end{bmatrix}, \tag{11}$$

где $S^0 = S$, $H^0 = H$, а $S^n = \sum_{k=0}^{K+n} s_k^n \delta_k$ и $H^n = \sum_{k=0}^{K+n} \sum_{j=m}^{M} s_k^n h_j^n \delta_k$ – функция источника и отклик от неё на $n$-том шаге итерации. При последовательном обнулении в функции источника $S^n$ коэффициенты $s_1^n = s_2^n = \cdots = s_n^n = 0$, поэтому:

$$S^n = s_0 \delta_0 + \sum_{k=n+1}^{K+n} s_k^n \delta_k, \quad H^n = s_0 \sum_{m=M_0}^{M} h_m \delta_m + \sum_{k=n+1}^{K+n} \sum_{m=M_0}^{M} s_k^n h_m^n \delta_{m+k}. \tag{12}$$

При числе итераций $N > K + M - M_0$ вторые слагаемые в (12) выйдут за пределы рассматриваемых областей определения функций $S$ и $H$, и в интервале $[0, K]$: $S^N \mapsto S_{opt} = s_0 \delta_0$, а в интервале $[M_0, K+M]$: $H^N \mapsto H_{opt} = s_0 \sum_{m=M_0}^{M} h_m \delta_m$, что означает решение задачи реконструкции.

На Рисунке 2 проиллюстрирован процесс реконструкции отклика $H$ от источника $S = S^0$ методом пошагового сдвига. Дискретность функции источника составляла 4 минимальных временных интервала, поэтому шаг итераций также составлял 4 интервала. Вертикальными штрих-пунктирными линиями обозначены рассматриваемые области определения функции источника ($K \approx 30$) и отклика ($K + M - M_0 \approx 60$). В этих интервалах функция источника принимает оптимальный вид на 8-ом шаге итерации ($8 \cdot 4 = 32 > K \approx 30$), а отклик – на 16-том ($16 \cdot 4 = 64 > K + M - M_0 \approx 60$). Таким образом, в данном примере $N = 16$ – минимальное необходимое количество итераций для реконструкции.

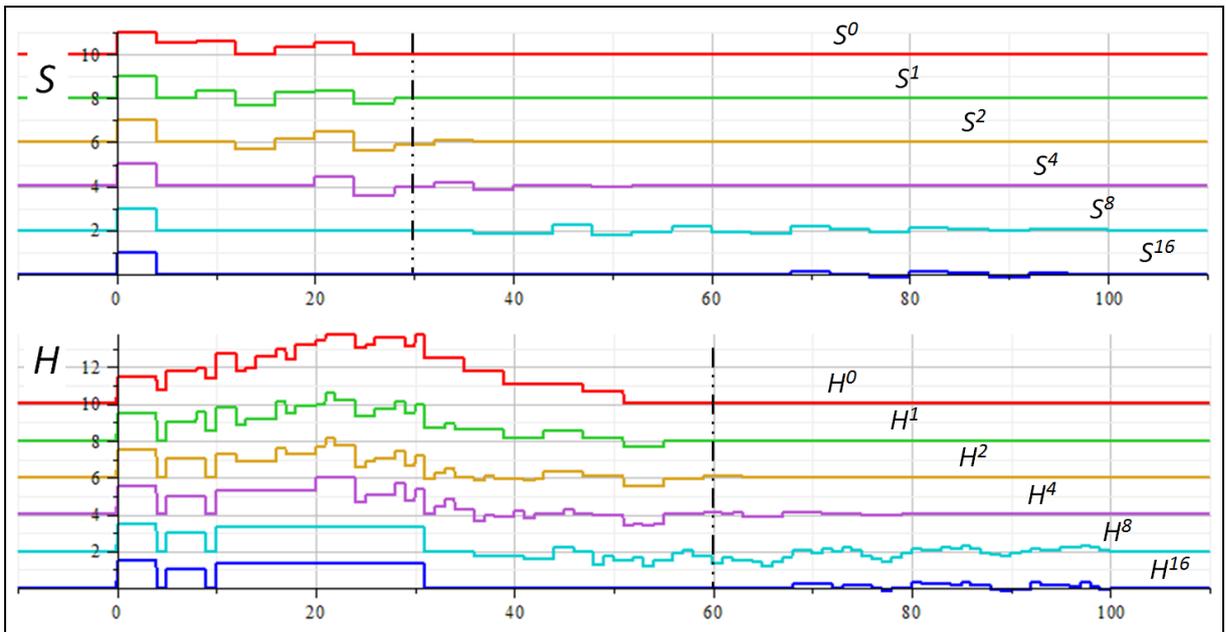

**Рисунок 2.** Реконструкция отклика методом пошагового сдвига.

На практике количество временных каналов значительно превышает количество каналов, рассмотренных в примере. Так, например, как указывалось ранее, общее количество временных каналов диагностического оборудования на установке ТРОНС достигает 149000. Несмотря на то, что метод реконструкции (5b) позволяет рассматривать отклик по частям, может потребоваться значительное количество итераций. Однако, в важных частных случаях метод пошагового сдвига можно модифицировать, значительно сократив требуемое число итераций, необходимых для реконструкции.

## 4. МОДИФИЦИРОВАННЫЙ МЕТОД

Пусть функция источника $S$ и чистый отклик $H$ от неё представлены в виде:

$$S = s_0 \delta_0 + s_k \delta_k,\ H = s_0 \sum_{m=M_0}^{M} h_m \delta_m + s_k \sum_{m=M_0}^{M} h_m \delta_{k+m}, \qquad (13)$$

то есть функция источника $S$ представляет собой два, разнесённых по времени, инициирующих импульса – основной с амплитудой $s_0$ и после-импульс с амплитудой $s_k < s_0$. Коэффициенты $s_1 = s_2 = \cdots = s_{k-1} = 0$, поэтому первые $k-1$ шагов итерационного процесса (11) не меняют ни функцию источника, ни отклик, и могут быть опущены. Первое изменение функций происходит на $k$-том шаге итерации (11), при этом у функции $S$ обнуляется коэффициент при $u_k$ и становится отличным от нуля коэффициент при $\delta_{2k}$. Следующие $k-1$ шагов итерационного процесса (11) также не изменят вид функций $S$ и $H$, которые будут меняться на $k$-том, $2k$-том, $4k$-том, и т.д. шаге итерации (11). Опуская в (11) шаги, не меняющие функций $S$ и $H$, и обозначая $a = s_k/s_0$, получим следующие итерационные уравнения:

$$S^{n+1} = S^n + a^{2^n} S^n_{2^n k},\ H^{n+1} = H^n + a^{2^n} H^n_{2^n k}, \qquad (14)$$

где $S^0 = S$, $S^1 = S^0 - a S^0_k$, $H^0 = H$, $H^1 = H^0 - a H^0_k$.

Модифицированный метод значительно сокращает число итераций, необходимых для реконструкции. Так, если $N$ – число итераций, необходимых для реконструкции методом пошагового сдвига, то число итераций $N^*$ модифицированным методом составит:

$$N^* = \log_2\left(\frac{N}{k}\right).$$

Например, даже при $k=1$, вместо 64 итераций, необходимых для реконструкции методом пошагового сдвига, потребуется всего 6 итераций модифицированным, вместо 128 – 7 итераций, вместо 256 – 8 и т.д.

Кроме того, если $a<1$, то амплитуда после-импульса в $S^n$ уменьшается в $a^{2^n}$ раз, и согласно (14) на столько же сокращаются искажения отклика $H^n$. В Таблице 1 представлены искажения отклика основного импульса в процентах от соотношения амплитуд импульсов $a<1$ и числа итераций $n$ модифицированным методом. Так, при $a=0.3$ погрешность от после-импульса составляет 30%, но уже после второй итерации погрешность составит всего 0.8%.

**Таблица 1.** Искажения отклика основного импульса (%) при $n$ итерациях.

| $a \setminus n$ | 0 | 1 | 2 | 3 | 4 | 5 |
|---|---|---|---|---|---|---|
| 0.9 | 90 | 81 | 65.61 | 43.05 | 18.53 | 3.434 |
| 0.7 | 70 | 49 | 24.01 | 5.77 | 0.332 | 0.001 |
| 0.5 | 50 | 25 | 6.25 | 0.391 | 0.002 | $2.33 \cdot 10^{-8}$ |
| 0.3 | 30 | 9 | 0.81 | $6.56 \cdot 10^{-3}$ | $4.30 \cdot 10^{-7}$ | $1.85 \cdot 10^{-15}$ |
| 0.1 | 10 | 1 | $10^{-2}$ | $10^{-6}$ | $10^{-14}$ | $10^{-30}$ |

На Рисунке 3 представлены диаграммы реконструкции отклика при функции источника $S = \delta_{0 \div 3} + 0.85 \cdot \delta_{8 \div 11}$. Достаточно трёх итераций, чтобы реконструировать отклик.

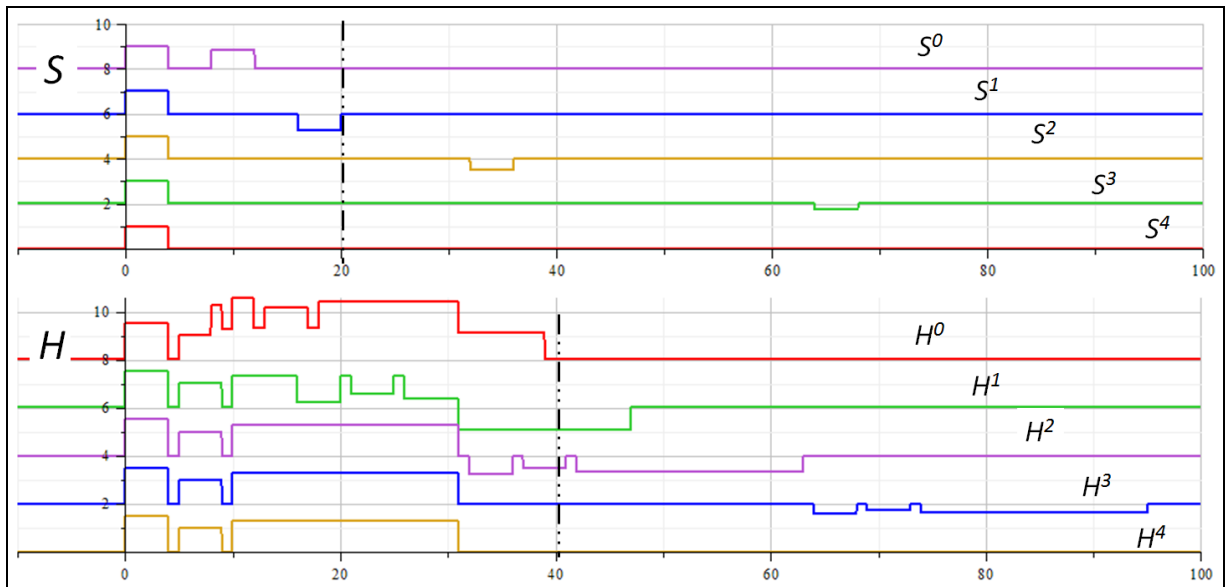

**Рисунок 3.** Реконструкция отклика модифицированным методом.

В случае $a=1$ уменьшения амплитуды искажений не происходит, и реконструкция осуществляется исключительно путём вывода искажений за области определения.

В случае $a>1$ применение как метода пошагового сдвига, так и модифицированного метода приведёт к экспоненциальному росту искажений и к расходимости итерационной последовательности при реконструкции. Задачу можно решить, если моделировать импульс $S$ из (13) в виде:

$$S = s_0 \delta_0 + \sum_p \frac{s_k}{p} \delta_k, \qquad (15)$$

где число $p$ выбирается таким, чтобы $a = s_k/(ps_0) \leq 1$. Последовательно устраняя влияние $p$ после-импульсов с амплитудой $s_k/p$ модифицированным методом, можно решить задачу реконструкции.

Модифицированный метод реконструкции (14) и принцип моделирования импульса (15) позволяют решить задачу реконструкции в общем виде для инициирующего импульса произвольной формы, в частности, для произвольного числа после-импульсов с произвольной амплитудой.

## 5. ПОВЫШЕНИЕ РАЗРЕШАЮЩЕЙ СПОСОБНОСТИ

Возможность устранения влияния выбранных частей инициирующего импульса является важным достоинством модифицированного метода. Эту возможность можно использовать для реконструкции отклика спектрометра от инициирующего импульса меньшей длительности или с более короткими фронтами, что будет соответствовать повышению разрешающей способности спектрометра, превышающей его конструктивные параметры вплоть до физических ограничений.

Пусть функция источника $S$ и чистый отклик $H$ от него представлены в виде:

$$S = s_0\delta_0 + s_0\delta_1, \quad H = s_0 \sum_{m=M_0}^{M} h_m \delta_m + s_0 \sum_{m=M_0}^{M} h_m \delta_{m+1}, \qquad (16)$$

то есть функция источника $S$ представлена в виде двух последовательных инициирующих импульсов вдвое меньшей длительности.

Уравнение (16) является частным случаем (13) при $s_k = s_0$ и $\delta_k = \delta_1$ и к нему применим модифицированный метод реконструкции, при которой, по существу, будет получен отклик спектрометра от инициирующего импульса вдвое меньшей длительности, что означает повышение разрешающей способности спектрометра.

На Рисунке 4 представлены диаграммы реконструкции отклика от импульса меньшей длительности при представлении функции источника $S = \delta_{0 \div 7}$ в виде $S = \delta_{0 \div 3} + \delta_{4 \div 7}$. Поскольку коэффициент $a = s_1/s_0 = 1$, то уменьшения амплитуды искажений не происходит и реконструкция осуществляется исключительно путём вывода искажений за область определения.

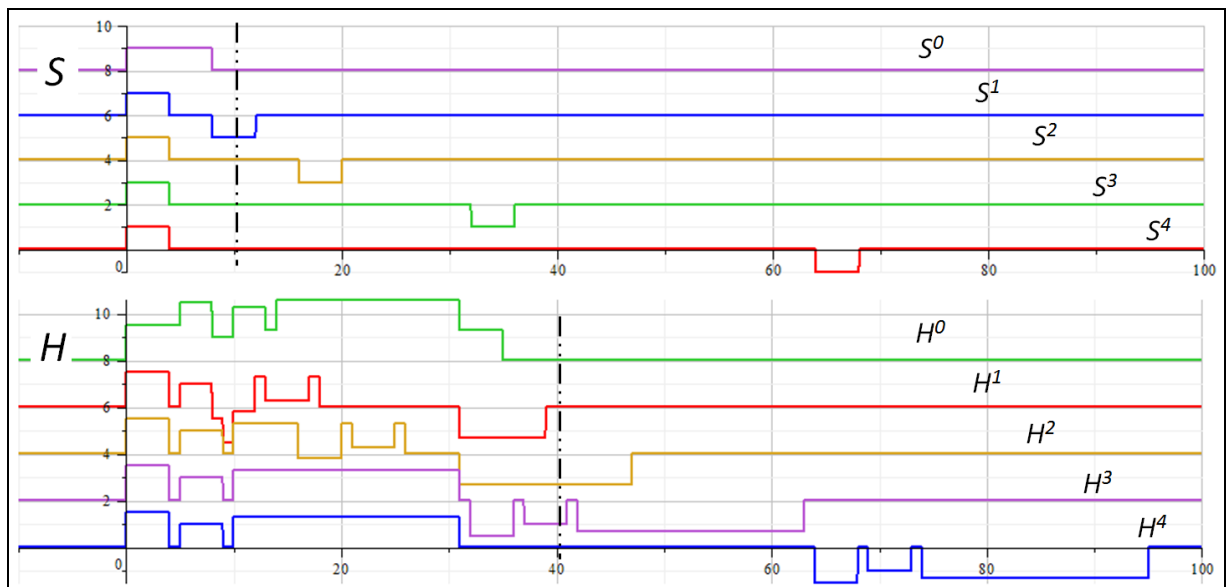

**Рисунок 4.** Повышение разрешающей способности модифицированным методом.

Аналогичным образом можно реконструировать отклик от импульса ещё меньшей длительности. Ограничивают возможность уменьшения длительности интервал дискретизации и физические процессы при образовании нейтронного потока и взаимодействия нейтронов с ядрами мишени.

Ещё одним ограничивающим фактором является наличие переднего фронта инициирующего импульса, при котором $s_0 < s_1 < \cdots < s_{\max}$, поэтому коэффициент $a = s_k/s_0 > 1$, что приведёт к расходимости итерационной последовательности. Задачу можно решить, применяя метод моделирования импульса (15), но при этом может понадобиться большое число итерационных последовательностей. Сократить число итерационных последовательностей можно методом моделирования фронтов.

Так, для реконструкции отклика при функции источника $S = \delta_{0 \div 1} + 2\delta_{2 \div 3} + \delta_{4 \div 5}$ (см. Рисунок 5) с помощью (15) потребуется как минимум три итерационных последовательности. Однако, представляя эту функцию в виде $S = \delta_{0 \div 3} + \delta_{2 \div 5}$, реконструировать отклик можно с помощью одной итерационной последовательности, что представлено на Рисунке 5.

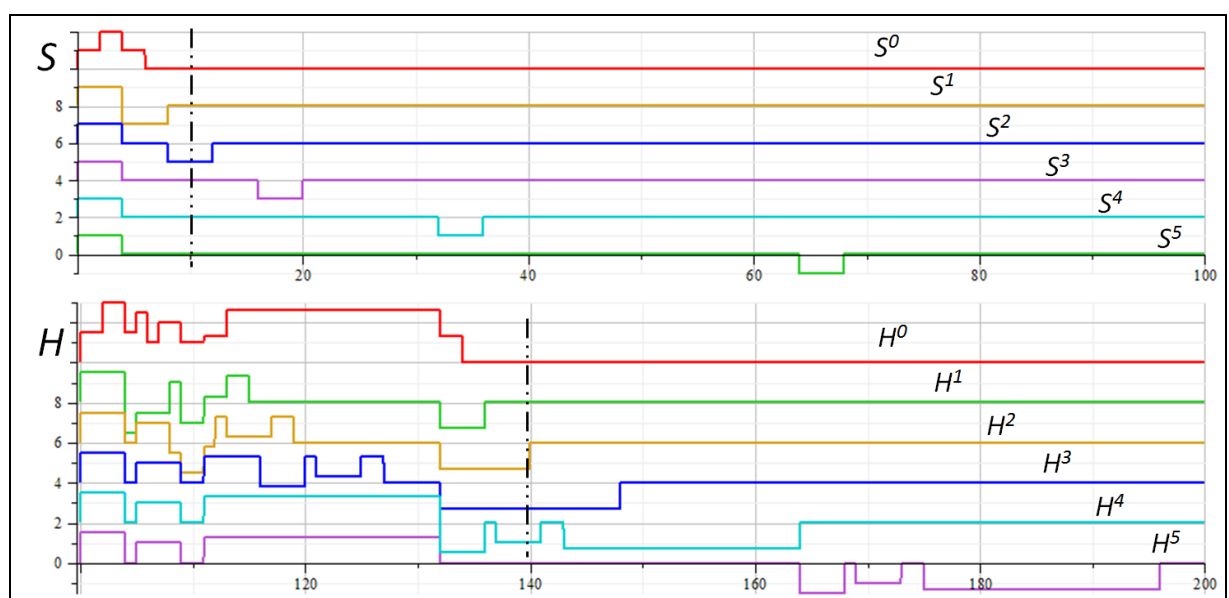

**Рисунок 5.** Реконструкция отклика импульса с передним фронтом.

Поскольку коэффициент $a = s_1/s_0 = 1$, то уменьшения амплитуды искажений не происходит и реконструкция осуществляется путём вывода искажений за область определения. Для сокращения числа итераций может оказаться полезным переход к сходимым итерационным последовательностям, например, путём представления рассмотренной функции $S = \delta_{0 \div 3} + \delta_{2 \div 5}$ в виде $S = \delta_{0 \div 3} + \frac{1}{2}\delta_{2 \div 5} + \frac{1}{2}\delta_{2 \div 5}$. В этом случае потребуется две итерационных последовательности, но в них, согласно Таблице 1, всего три шага итерации, чтобы достигнуть точности 0.4%.

Таким образом, функция с фронтом может быть приведена к прямоугольному виду с помощью представления функции источника в виде:
$$S = \sum_k a_k U_k, \text{ где } U_k = \delta_{k \div k+m} \text{ и } |a_k| \leq 1. \tag{17}$$

На практике, при сложной форме инициирующего импульса, возможно усреднить его до прямоугольника с длительностью, определяемой по полувысоте импульса. В этом случае точность реконструкции будет определяться этим усреднением.

На Рисунке 6 представлены диаграммы реконструкции отклика спектрометра модифицированным методом с усреднением коэффициента преобразования $a$ по полувысоте инициирующего импульса в области резонансов, показанных на Рисунке 1 в серии измерений 2. Реконструировалась область длительностью 400 каналов, что составляет 150 нс·400=60 мкс. Значение коэффициента преобразования составило $a = 0.5$, шаг итерации определялся исходя из задержки между основным и после-импульсом и составил 6 мкс. Трёх итераций оказалось достаточно для устранения ложных пиков, вызванных после-импульсом в серии 2.

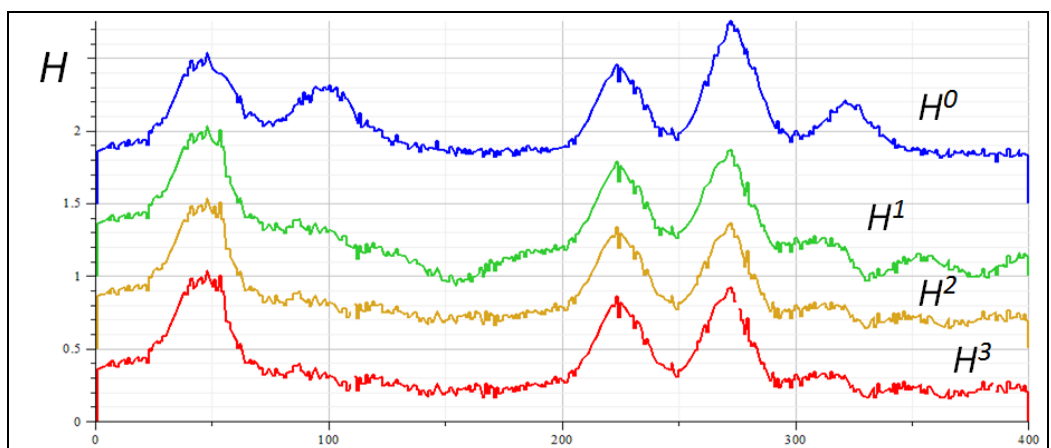

**Рисунок 6.** Реконструкция отклика спектрометра ИЯИ РАН (Ta-181).

Повышение разрешающей способности спектрометров по времени замедления сопряжено с большими техническими трудностями, такими как уменьшение длительности инициирующего импульса или увеличение пролётной базы. Так, например, для увеличения разрешающей способности в два раза требуется примерно вдвое увеличить пролётную базу (в ТРОНС ИЯИ РАН с 50 до 100 метров), при этом интенсивность потока нейтронов упадёт примерно в четыре раза. Повышение разрешающей способности спектрометра математическими методами не требует больших затрат и является весьма перспективной возможностью.

## ЗАКЛЮЧЕНИЕ

Предлагаемые математические методы реконструкции откликов времяпролётных спектрометров позволяют путём обработки экспериментальных данных устранить искажения в них, вызванные нестабильностью инициирующих импульсов, и оптимизировать измерения. Эти методы также позволяют в определённых рамках повышать разрешающую способность существующих спектрометров, что является перспективной альтернативой техническим методам, сопряжённым с трудноразрешимыми научно-техническими проблемами и требующим больших материальных затрат.

## СПИСОК ЛИТЕРАТУРЫ